# Analysis of Drone-Assisted Building Inspection Training in VR vs 2D Monitor Display: an EEG Study

Pengkun Liu, Ph.D.[1]; Jackson Greene [2]; Jiali Huang, Ph.D.[3];
Pingbo Tang, Ph.D. P.E.[1]; and Yu Hou, Ph.D.[2*]

[1]Carnegie Mellon University, Department of Civil and Environmental Engineering, 5000 Forbes Ave, Pittsburgh, PA 15213; e-mail: pengkunl@andrew.cmu.edu, ptang@andrew.cmu.edu
[2]Western New England University, Department of Construction Management, 1215 Wilbraham Rd, Springfield, MA 01119; e-mail: jackson.greene@wne.edu, yu.hou@wne.edu
[3]Western New England University, Department of Industrial Engineering and Engineering Management, 1215 Wilbraham Rd, Springfield, MA 01119; e-mail: jiali.huang@wne.edu
[*]corresponding author. ORCID: 0000-0002-9822-244X

## ABSTRACT

Researchers have been using simulation-based methods for drone-assisted inspection training. Multiple brain regions are associated with information processes and decision-making, and the connectivity of these regions may further influence inspectors' performance. However, researchers do not understand the pathways of the information flows when drone pilots process the maintenance and manipulation of information, which may affect the efficiency of tacit knowledge transfer. This study aims to reveal the causal connection between participants' brain regions using an electroencephalogram and dynamic causal modeling when processing drone-assisted building energy audit tasks using different display modalities. The results showed similar single-direction connectivity patterns for the different simulation groups. The results also showed similar patterns between brain regions related to visual inspection performance before and after training. These findings highlight the nature of brain asymmetries and may be utilized in measuring cognitive states and designing adaptive automation in the knowledge transfer of drone-based inspection.

## INTRODUCTION

Innovative drone technologies have made building inspections and energy audits more intelligent. For example, researchers have installed multiple sensors on drones, including infrared thermal cameras, to detect heat loss, abnormal temperature, thermal bridges, and moisture areas in building envelopes. Based on the results of these inspections and analyses, homeowners and facility managers can take retrofitting actions to improve building energy efficiency.

Autonomous systems have been applied to drone-assisted inspection tasks, but manual mode is still critical since drone pilots must use their previous experience to inspect certain checkpoints that are vulnerable to building degradation and locations in which heat loss occurs frequently. This experience is defined as tacit knowledge and requires trainees to obtain it by practicing. In addition, it requires pilots to use short- and long-term memory to process information and make decisions. For example, short-term memory refers to remembering which locations have been inspected, while long-term memory refers to recalling the inspection tasks' initial conditions and general criteria. Trainees who obtain skills from senior engineers can efficiently and accurately audit building facade energy performance. Therefore, providing such workforce training for trainees and transferring tacit knowledge from senior engineers to them is important.

Researchers have been using simulation-based methods for drone-assisted inspection training, such as Virtual Reality (VR) and Augmented Reality (AR). However, current workforce training cannot attract the younger generations' interest. The potential reasons could be ineffective interaction with explainable operational strategies, lack of theoretical preparation, and insufficient interactive demonstration that provides timely responses. In addition, researchers also want to know the correlation between the performance of the younger generation workforce and their brain activities. With understanding such correlation, researchers can examine the pathways of the information flows when drone pilots process the information maintenance and manipulation, which may be critical factors affecting the efficiency of tacit knowledge transfer. Multiple brain regions have been shown to be associated with the cognitive processes involved in drone control, such as the primary visual cortex (V1), which is responsible for visual awareness. , the fusiform gyrus, contributing to color information processing (Mueller et al. 2012), and the dorsolateral prefrontal cortex (dlPFC), which is for executive functions, cognitive control, and processing emotions (Friedman and Robbins 2022).

This study examines connections among various brain regions and explores the relationships between brain activities and trainees' inspection performance. To investigate the influences of different simulation-based methods, we chose two display modalities: VR and 2D monitors. The participants were divided into two groups and asked to complete an energy audit task of the building. After receiving training on energy audits, they returned to a simulated environment and redid the experiments. We hypothesize that (1) the right-lateralized and backward-only connection patterns of trainees' brain regions will be observed for the VR-based simulation group. In contrast, bi-directional, widespread connection patterns will be observed in the monitor-based simulation group. (2) The connectivity patterns among V1-, Fusiform gyrus-, and dlPFC-related connection networks will vary before and after energy audit-related training.

**RELATED WORK**

**Drone-Based Built Environment Inspection**

Drone-based inspection has emerged as a transformative approach for assessing and monitoring buildings and infrastructure. Utilizing unmanned aerial vehicles (UAVs) or drones, this method offers significant advantages over traditional inspection techniques (Bolick et al. 2022). Equipped with a variety of sensors and cameras—including thermal, infrared (Hou et al. 2022), visible light, multispectral, and hyperspectral—drones can capture diverse data types. Drone-based inspections involve dynamic tasks requiring operators to manage multiple demands simultaneously. These include drone piloting, interpreting sensor data, adhering to safety protocols, and making critical decisions. Navigating varied inspection environments, such as bridges, industrial facilities, and high-rise buildings, compounds these cognitive demands. Effective management of this cognitive load is crucial to ensuring operator efficiency, safety, and the accuracy of inspection outcomes.

**Simulation-Based Workforce Training and Performance Assessment with Electroencephalogram (EEG)**

Immersive learning is an educational approach that utilizes advanced technologies to simulate real-world environments, enabling learners to engage deeply with content in dynamic and interactive ways. By harnessing immersive technologies such as VR, AR, and Mixed Reality (MR), this method creates experiences that surpass traditional classroom or online education. Immersive simulation-based learning allows learners to practice their skills and knowledge safely and effectively by creating realistic computer simulation scenarios. Researchers have developed virtual environments in the construction industry to train engineers in inspecting highway construction and bridges (Liu et al. 2023). Immersive simulation-based training can be a crucial means of transferring knowledge, as it captures actions and movements, enabling trainees to observe and experience tasks in high fidelity (Makransky and Petersen 2021). This approach can be scaled to train multiple individuals simultaneously, reducing the training burden on senior inspectors. However, due to the complexity of inspection tasks, especially those involving drones, simulation-based training for such activities has not been sufficiently investigated to improve knowledge transfer efficiency and learning comfort. Drone-based inspections involve complex and dynamic interactions among humans, drones, and environments, posing significant challenges for training and assessment. Therefore, there is a need to develop and evaluate simulation-based training methods that can effectively train and measure drone operators' inspection skills and strategies.

Researchers employ physiological data (quantitative) and self-report measures (qualitative) to evaluate participants' performance during immersive training. Physiological data, including neuropsychological signals such as electroencephalography (EEG) for brain activity, as well as electrocardiograms (ECG), eye tracking, skin temperature, and thoracic posture, are collected using wearable devices (Sakib et al., 2021). EEG technology provides a robust method for quantifying cognitive load during drone operations and training programs. EEG measures brain activities, offering detailed insights into cognitive processes related to attention and working memory. Research has identified key brain regions involved in working memory processes, such as the dorsolateral prefrontal cortex (dlPFC) and parietal cortex, which show increased activation under high cognitive load. EEG data can reveal how operators process and manage tasks during drone-based inspections by analyzing oscillatory activity and functional connectivity within these regions. Integrating EEG into simulation-based training environments can enhance learning outcomes by personalizing training and fostering a deeper understanding of cognitive processes.

**Dynamic Causal Modeling**

While traditional EEG studies could reveal the temporal changes within a pre-determined power band (e.g., Chikhi et al. 2022), we want to dig deeper in terms of the interplay of multiple brain regions. Dynamic Causal Modeling (DCM) serves as a useful tool to infer effective connectivity, which is the directional causal relationship between brain regions. First introduced in 2003 for functional Magnetic Resonance Imaging (fMRI), DCM has gained interest in the field of neuroimaging to reveal the information pathways among brain networks (Friston et al. 2003). It was later adapted for EEG signals (Kiebel et al. 2008) and utilized to study the connections underlying cognitive processes such as mental workload, emotion, working memory, and motor imagery. DCM utilizes a process named Bayesian Model Selection (BMS) to determine the most likely connection model for the observed data (e.g., EEG recordings). The exceedance probability is calculated for the proposed candidate models, denoting how much one model is more likely to generate the observed data than the others (Stephan et al. 2009). The winning model describes the causal relationship (i.e., effective connectivity) and provides evidence for better working environment designs.

Using DCM for the EEG method, we could shed some light on the information flow during the heat loss detection in a simulated environment. Former neuroimaging studies identified brain regions activated in the occipital cortex, parietal cortex, and frontal cortex during tasks involving drone control (e.g., Khan and Hong 2017). Regions in the temporal cortex (e.g., Fusiform gyrus) were also found to be activated when virtual reality is involved (Mueller et al. 2012). However, not much information is found concerning the direction and the laterality of the connections between those activated regions. By constructing candidate models and selecting the best-fitted model using DCM, we could gain a deeper understanding of this matter.

**METHODS AND EXPERIMENTS**

As shown in Figure 1, this research uses an EEG cap, VR goggles, and monitors to conduct experiments, collect, and process EEG signals, and build the relationships between participants' brain region connectivity (such as nine models in the figure) and their performance on a drone-assisted building inspection. Participants were divided into two groups: VR and monitor modalities. The BMS shows two different winning models for these two modalities.

**Participants**

A total of 4 participants (all male, average age = 21.8, SD = 2.67) from a local university were recruited. All participants completed the entire experiment, and their data were analyzed. All participants reported being free of medical or neurological disorders and had normal or corrected vision. Written consent was obtained from all participants. The experiments were reviewed and approved by the University's Institutional Review Board. Participants were divided into two groups: VR- and 2D monitor-based. The ID numbers of participants were P-001 and P-003 in the VR-based group and P-002 and P-004 in the 2D monitor-based group.

**Major Apparatus**

- **EEG:** a 32-channel EEG system (EPOC Flex from Emotiv, San Francisco, CA)
- **VR Goggle:** Meta Quest 3 (Meta, Menlo Park, CA)
- **Monitor:** a 24 Monitor with a resolution of 1920 x 1080 and a panel using In-Plane Switching (IPS)(P2425 from Dell, Round Rock, TX)

**Experimental Procedure**

**Step 1 Pre-experiment:** The experimental procedure began by obtaining informed consent from each participant, acknowledging their voluntary participation. Following this, participants completed a questionnaire designed to document their prior knowledge of VR and heat loss detection, as well as any visual conditions that might influence the experiment's outcomes. This initial step ensured that any baseline factors related to the participants' experience and visual acuity were accounted for.

**Step 2 Applying Apparatus:** We instructed participants to sit and carefully put an EEG cap on their heads. The conductive gel was applied to ensure a high-quality signal, and 32 EEG electrodes were positioned to establish direct contact with the subject's scalp. With the EEG cap in position, we were ready to monitor brain activity for the following drone simulation task. After the EEG setup, participants in the VR-based group were directed to stand and put on the VR headset, which immersed them into a virtual environment where they would operate a drone, while participants in the 2D monitor-based group operated a drone by watching the virtual environment through the monitor screen.

**Step 3 Experiment phase:** Once placed in the virtual environment, participants were given a controller and instructed to fly the drone along the building facades, capturing images of heat loss. A brief tutorial was given to familiarize participants with the drone's controls. Before learning the related knowledge of building envelope energy audits, participants were given 3 minutes to fly the drone and identify as many heat loss instances as possible based on their prior knowledge, refraining from interaction except to signal the end of the allotted time. The next phase involved a training section. Participants read a detailed document explaining the science of heat loss, which included an overview of the different types of heat loss that can occur within buildings, to train the participants to successfully detect all heat loss from buildings. Following this training, we guided the participants in flying the drone back along building facades in the virtual environment and asked them to re-inspect the building facades to find heat loss, which was also within 3 minutes.

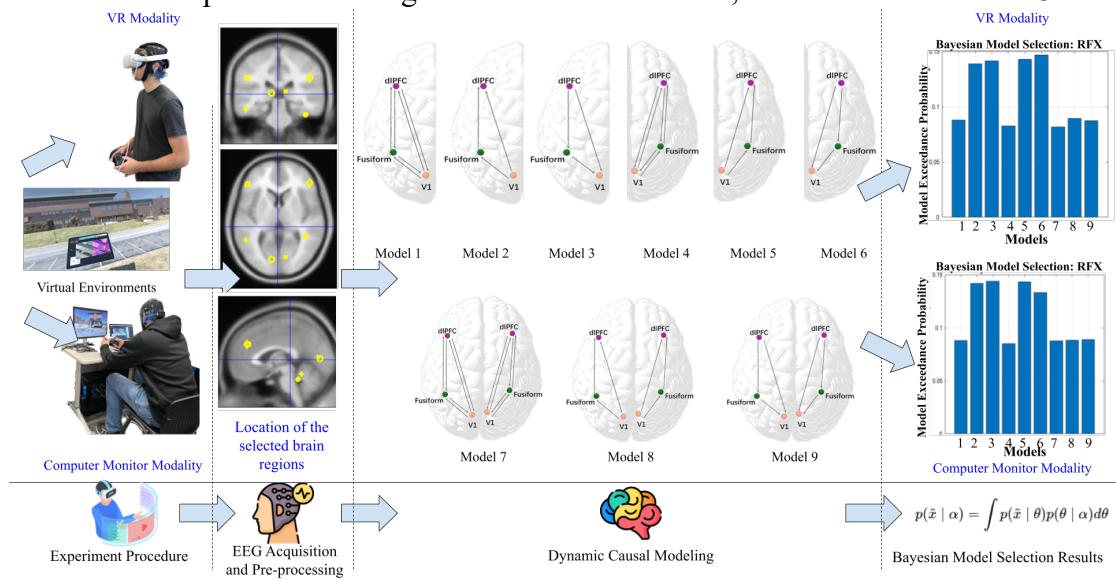

**Fig. 1.** Framework of Research Methods and Experiment Procedures.

**EEG Acquisition and Pre-processing**

EEG recordings were acquired from a 32-channel EEG system, sampled at 256 Hz. The 32 electrodes were arranged according to the 10-20 system (Sharbrough et al. 1991). Recordings were referenced to the ear lobes. The EEG signal pre-processing was conducted using MATLAB (The

MathWorks), and SPM 12 (Penny et al. 2011). The recorded data were band-pass filtered between 0.5 and 50 Hz to take out unwanted frequency bands. The continuous 3-min EEG data were segmented into a set of 2-s epochs without overlap and averaged.

**Dynamic Causal Modeling**

The pre-processed data were used to infer the causal connections underlying the experimental conditions. To construct the model space, we selected the bilateral dlPFC, fusiform, and primary visual cortex (V1). These brain regions were selected based on previous VR and drone control neuroimaging research and brain network studies (Kiebel et al. 2006). The coordinates and the locations in the MNI space can be found in Table 1 and Figure 1.

To infer the connectivity, each source was modeled as a single equivalent current dipole (ECD) to map the signals from sensors to sources (Kiebel et al. 2006). A total of nine connection models were constructed among the six brain regions to account for connection direction as well as laterality. The model space can be found in Figure 1.

Table 1. Coordinates of the nodes in MNI space.

| ROIs | l-V1 | r-V1 | l-Fusiform | r-Fusiform | l-dlPFC | r-dlPFC |
|------|------|------|------------|------------|---------|---------|
| x    | -11  | 11   | -51        | 39         | -48     | 48      |
| y    | -81  | -78  | -51        | -51        | 36      | 38      |
| z    | 7    | 9    | -18        | -24        | 30      | 30      |

**RESULTS AND DISCUSSION**

**Participants' Performance**

The participants' performances in finding abnormal areas from building facades are summarized in Table 2. The criteria for assessing participants' performance are Recall, Precision, and F1 values. In this case study, (1) participants in the 2D Monitor team performed better than the VR team. The cost of learning with different display modalities could influence the performance. The participants may spend more time to be familiar with the VR environments. (2) The participants in both teams improved their performance after training. It should be noted that the performance gaps between the VR team and the Monitor team were narrowed before and after training. This indicates that the influence of learning costs with different display modalities is weakened when participants are familiar with VR environments. The participants' flight paths are visualized in Table 3. The dark purple color represents their drones' starting points, and the light yellow color represents the drones' stop points. Most participants' drone flights followed a certain planned pattern and were regular. For example, P-001 and P-004 flew the drone from top to bottom and right to left. However, P-003's flight path was chaotic. This participant randomly operated the drone to inspect the building facade, although P-003's performance was not the worst.

**Brain Connectivity**

The results of the model selection are shown in Table 3. For both VR and monitor simulation scenarios, the models with the highest exceedance probability were model 2, model 3, model 5, and model 6, regardless of training conditions for all participants. The only exception is the post-training monitor-based simulation condition for the P-004. We observed the strongest evidence for model 7 in this scenario.

Models 2, 3, 5, and 6 are all patterns with unidirectional and unilateral connections. Forward and backward connections were almost equally likely to occur in the building inspection scenarios, but they do not co-exist simultaneously. Left-lateralized connections and right-lateralized connections were also equally likely to occur, yet they seldom manifest at the same time (the P-004's post-training scenario is the sole exception).

Past connectivity studies investigated the direction and the laterality of the connection pattern in various cognitive processes. Relevant examples include mental workload and training. Our past study on mental workload showed that the causal connections shifted from the left to both sides of the brain with increased workload (Huang et al. 2024). This was not observed in the present study, possibly due to the nature of the task and the experiment design. Participants were not given a task that demanded their full processing capacity, thus only unilateral patterns were observed. In terms of the effect of training, while an improvement in performance was observed for both the VR team and the Monitor team after training, the BMS reported similar winning models. Prior studies that focused on prolonged training periods observed altered connectivity patterns after the training (e.g., Sun et al. 2014). Further investigations are needed to clarify the relationship between training time and neural plasticity.

**Table 2. Participants' Performance.**

| Experiments | Results | VR | | Monitor | |
|---|---|---|---|---|---|
| | | P-001 | P-003 | P-002 | P-004 |
| Before Training | Recall | 23.08% | 23.08% | 53.85% | 76.92% |
| | Precision | 50.00% | 75.00% | 100.00% | 100.00% |
| | F1 | 31.58% | 35.29% | 70.00% | 86.96% |
| After Training | Recall | 53.85% | 61.54% | 61.54% | 84.62% |
| | Precision | 100.00% | 100.00% | 72.73% | 91.67% |
| | F1 | 70.00% | 76.19% | 66.67% | 88.00% |

*Notes:* $\text{Recall} = \frac{TP}{TP+FN}$   $\text{Precision} = \frac{TP}{TP+FP}$

***True Positive (TP):*** Participants correctly classified positive instances (e.g., abnormal areas).
***False Negative (FN):*** Participants incorrectly classified negative instances. (e.g., taking a picture of normal areas).
***TP + False Positive (FP):*** The ground truth of all abnormal areas.

**Research Limitations**

**Sample Size:** Considering the complexity of experiment configurations, we only had four participants. This project aims to explore the feasibility of the experiments and summarize the preliminary results. We expected to recruit more participants in the future to generalize our findings. For example, more evidence will explain the influences of learning costs with VR.

**Table 3. Participants' BMS Results.**

| Experiments | Results | Participants | | | |
|---|---|---|---|---|---|
| | | VR | | Monitor | |
| | | P-001 | P-003 | P-002 | P-004 |
| Before Training | Flight Path | 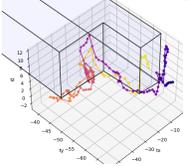 | 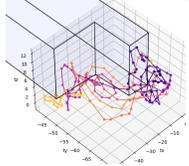 | 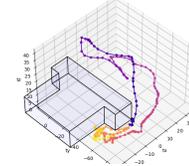 | 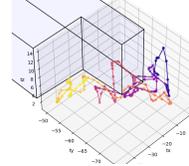 |
| | Bayesian Model Selection | 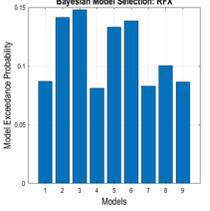 | 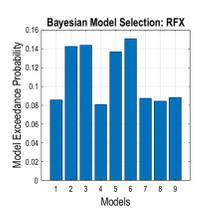 | 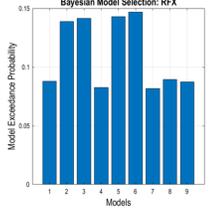 | 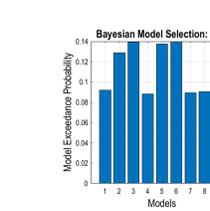 |
| After Training | Flight Path | 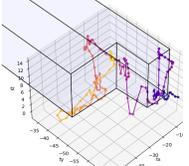 | 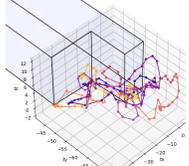 | 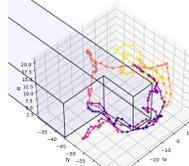 | 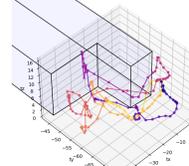 |
| | Bayesian Model Selection | 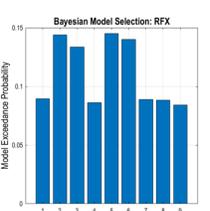 | 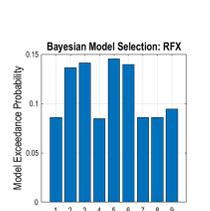 | 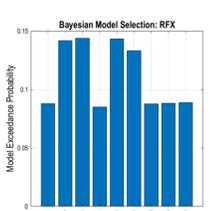 | 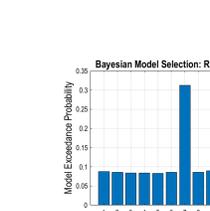 |

**Comprehensive Brain Region Connectivity:** To fully understand the effect of the simulation environment and training time on the brain connection patterns, a more comprehensive model space is needed. More brain regions should be included as connectivity nodes to reveal the full picture of the information flow.

**Users' Experiences:** Physical discomfort and fatigue could interfere with participants' task performance using EEG devices. To ensure the quality of the EEG signal collection, participants were asked to stay still and adjust the electrodes' positions on their scalp before starting each session. Although participants were not interfered with during the experiments, the tedious and complicated preparation may have influenced their mood to finish each experiment.

**CONCLUSION**

This study examines the connections among various brain regions, including V1, Fusiform gyrus, and dlPFC regions, and explores the participants' inspection performance. We chose two display modalities. Participants were asked to inspect building facades and find heat loss using a simulated drone in a virtual environment. They used their prior knowledge to finish tasks and redo the experiments after a training session. This research revealed that (1) unidirectional and unilateral connections were found in the brain regions. Forward and backward connections were almost equally likely to occur in the VR and monitor-based teams. (2) The connectivity patterns vary slightly before and after energy audit-related training. We will recruit more participants and add experiments with new and complicated inspection tasks as we analyze and discuss. More brain regions should be included as connectivity nodes to reveal the full picture of the information flow.